\documentclass[namedreferences]{solarphysics}
\usepackage[optionalrh]{spr-sola-addons} % For Solar Physics 
\usepackage[pdfborder={0 0 0 },urlcolor=blue,breaklinks]{hyperref}
\ifx \doiurl \undefined \def \doiurl#1{\href{http://dx.doi.org/#1}{\url{#1}}}\fi
\ifx \adsurl \undefined \def \adsurl#1{\href{http://adsabs.harvard.edu/abs/#1}{\url{#1}}}\fi

\usepackage{graphicx}        % For eps figures, newer & more powerfull
\usepackage{color}           % For color text: \color command
\usepackage{url}             % For breaking URLs easily trough lines
\newcommand{\kms}{km~s$^{-1}$}

\newcommand{\aap}{    {\it Astron. Astrophys.}}

\newcommand{\apj}{    {\it Astrophys. J.}}
\newcommand{\apjl}{   {\it Astrophys. J. Lett.}}

\newcommand{\solphys}{{\it Solar Phys.}}

\begin{document}

\begin{article}

\begin{opening}

\title{Velocities and Linewidths in the Network and Cell Interiors of a
Polar Coronal Hole, Compared with Quiet Sun}

\author{K.P.~\surname{Raju}$^{1}$\sep
        B.J.I.~\surname{Bromage}$^{2}$
       }
\runningauthor{K.P. Raju, B.J.I. Bromage}
\runningtitle{Velocities and Linewidths in the Polar Coronal Hole Network}

   \institute{$^{1}$ Indian Institute of Astrophysics,
                     Bangalore-560034, India\\
                    email: \url{kpr@iiap.res.in} \\ 
              $^{2}$ Jeremiah Horrocks Institute, \\
                     University of Central Lancashire, Preston, PR1 2HE, UK\\
                     email: \url{bjibromage@uclan.ac.uk} \\
             }

\begin{abstract} The relative Doppler velocities and linewidths in a polar
coronal hole and the nearby quiet-Sun region have been  obtained from the {\it Solar 
and Heliospheric Observatory} (SOHO)/{\it Coronal Diagnostic Spectrometer} (CDS) observations 
using emission lines originating at different heights in the solar atmosphere from  the  lower 
transition region (TR) to the low solar corona.
The observed region is separated into the network and the cell interior and the 
behavior of the above parameters were examined in the different regions.
It has been found that the histograms of Doppler velocity and width are 
generally broader in the cell interior as compared to the network. 
The histograms of Doppler velocities of the network and cell interior
do not show significant difference in most cases. However, in the case of the 
quiet Sun, the Doppler velocities of the cell interior are more blueshifted than 
those of the network for the lowermost line He {\sc ii} 304 {\AA}, and an opposite
behavior is seen for the uppermost line Mg {\sc ix} 368 {\AA}.
The histograms of line width show that the network--cell difference is more prominent 
in the coronal hole. The network has significantly larger linewidth than the cell interior for the 
lowermost TR line He {\sc ii} 304 {\AA} for the quiet Sun. For coronal hole, this is true 
for the three lower TR lines He {\sc ii} 304 {\AA}, O {\sc iii} 599 {\AA}, 
and O {\sc v} 630 {\AA}. Also obtained are the correlations between the relative 
Doppler velocity and the width. 
A mild positive correlation is found for the lowermost transition
region line He {\sc ii} 304 {\AA} which further decreases or become
insignificant for the intermediate lines. For the low coronal 
line,  Mg {\sc ix} 368 {\AA}, the correlation becomes strongly negative.
This could be due to the presence of standing or propagating waves
from the lower to the upper solar atmosphere.
The results may have implications for
the generation of the fast solar wind and coronal heating.
\end{abstract}
\end{opening}
%-------------------------------------------------

\section{Introduction}
     \label{S-Introduction} 

The chromospheric network, the bright emission network seen in the 
chromospheric lines such as Ca {\sc ii} and H$\alpha$ lines, represents
the boundaries of the supergranulation cells \cite{Sim64}.
The convective motions in the cells sweep small flux tubes to the edges of the cells, 
resulting in magnetic concentrations and enhanced emission there.
The EUV emission network is essentially the continuation of the 
chromospheric network in the transition region \cite{Brue74,Reev74}.
The network slowly disintegrates and becomes indistinguishable from 
the surroundings in the corona.

The morphological properties and the evolution of the network are different in the 
quiet Sun and the coronal hole. {\it Skylab} observations in the 1970s have shown that 
the network emission is weakened in coronal holes as compared to quiet-Sun 
region \cite{Huber74}. \inlinecite{Gall98} have obtained
the intensity contrast of the network with respect to the internetwork for 
different lines in the quiet Sun, and the maximum was found for the
O {\sc v} 629.73 {\AA} line whose formation temperature is about 0.25 MK.
\inlinecite{Raju10} obtained the intensity contrast of the network for 
different lines in both quiet Sun and coronal hole. The contrast, 
in general, is lower for the coronal hole as compared to the quiet Sun, but 
becomes equal in the upper transition region. The maximum contrast for 
both the regions was found at about 0.2 MK. These results seem to suggest that the 
Doppler velocities and widths could also be different in the network and the 
cell interior and also there are possible variations between quiet Sun and coronal hole.

It has been known that there is a systematic difference between coronal hole and 
the quiet Sun in terms of intensity, velocity, and width \cite{Raju09}.
\inlinecite{Judge97} report that Transition region (TR) lines show more redshift in the network 
than in the inter-network and also there is a correlation between line intensity
brightenings and increased redshift.
\inlinecite{Gont01} also find that velocity distribution is different for 
network and inter-network with network having more redshift with lower standard 
deviation. However, \inlinecite{Hass99} find that plasma outflow in coronal
hole originates predominantly along the network boundaries.
\inlinecite{Pope04} report a correlation between network intensity and Doppler 
velocity for the O\,{\sc iii} line.

In the present article, spectroscopic data from  {\it Solar and Heliospheric Observatory} 
(SOHO)/{\it Coronal Diagnostic Spectrometer} (CDS) were used to examine the 
variations of Doppler velocities and emission linewidth in the network and
the cell interior in a polar coronal 
hole (PCH) and the quiet Sun region outside. Observations were made in five 
different emission lines whose formation temperatures vary from 0.08\,--\,0.95 MK
and hence represent the lower transition region to the  inner corona. 
In particular, two aspects have been examined; i) evolution of the network
in the solar atmosphere, and ii) the correlation between Doppler
velocity and linewidth in the different regions.  

\section{Data and Analysis}
  \label{S-Data} 

SOHO/CDS  observations   made  during  the  Whole  Sun  Month
(August -- September 1996) have been used.   Details  of  the  observations
 and data reduction have been given by \inlinecite{brom00} and \inlinecite{Raju09}.  
Only five strong lines  were selected for the present analysis. 
Details of the emission lines were obtained from CHIANTI \cite{Dere97,Land06} and 
may be seen in Table 1.
The CDS field of view is $240^{\prime\prime}  \times 60^{\prime\prime}$, which 
is a combination of three rasters. The   spatial resolution is approximately
$4^{\prime\prime}$ but a running averaging of five nearby pixels has reduced 
this to about $20^{\prime\prime}$ and the overall temporal resolution to about an hour.
14 datasets from  the north polar region near the central meridian are used in this
study.
The {\it Normal Incidence Spectrometer} (NIS) data suffer from instrumental trends 
due to the rotation and tilt effects \cite{Brooks06}. These are corrected as described by 
\inlinecite{Raju09}.

The spectra obtained from the CDS windows were fitted with single or multiple 
Gaussians  depending upon the number of lines present in the window using the routine
{\sf CFIT-BLOCK} in Solar Software (SSW). 
The individual line profiles were always fitted with a single Gaussian. The He 
window has two strong lines, He {\sc ii} at 304 {\AA} (second order at 607.56 {\AA})
and O {\sc iv} at 608.31 {\AA}, and hence a double Gaussian fit was used. The 
O {\sc iii} 599 {\AA} and O {\sc v} 630 {\AA} windows were fitted with single 
Gaussian. The Ne {\sc vi} window was fitted with a double Gaussian to fit 
Ne {\sc vii} 561.7 {\AA} and Ne {\sc vi} 562.8 {\AA}. The  Mg {\sc ix}  window
was fitted with a double Gaussian to fit Mg {\sc vii} 367.7 {\AA} and
Mg {\sc ix} 368 {\AA}. 
It may be noted that 
the zero point in velocity is kept equal to the most probable value in the 
data, and hence the Doppler velocities in the present analysis are relative
Doppler velocities with respect to the coronal rest frame. 
Only those line profiles with a signal-to-noise ratio greater than ten were
selected for the analysis. With this, the estimated errors are 2 {\kms} 
in velocity and  0.03 {\AA} in width.  

\begin{table}[h]
\caption{ Details of the emission-line parameters. Note that He line is observed in 
the second order.
}
\label{T-simple}
\begin{tabular}{clcl}    
  \hline                   % horizontal line
No.&Ion&$\lambda$&T\\
   &   &[\AA]    &[MK]\\
  \hline
1&He {\sc ii}&303.78&0.083\\
2&O {\sc iii}&599.59&0.11\\
3&O {\sc v}&629.73&0.25\\
4&Ne {\sc vi}&562.80&0.43\\
5&Mg {\sc ix}&368.06&0.95\\
  \hline
\end{tabular}
\end{table}

The coronal hole and the quiet Sun were identified in the images on the basis of 
the intensity of the Mg {\sc x} 625 {\AA} line. The network and the cell interior are 
defined on the basis of the intensity distribution of the O {\sc v} 629.73 {\AA} line.
Those points for which the intensity is above two-thirds of the distribution are taken 
as network. \inlinecite{Reev76} took the average intensity to distinguish
the network and the cell interior, whereas some authors \cite{Xia04} have taken the two-thirds
criterion. We find that the results remain the same irrespective of the two criteria. 

\section{Results}
\label{S-Results} 

Histograms of relative Doppler velocities in the network and the cell interior in the
quiet Sun and coronal hole in different emission lines are given in Figure~1. 
 The bin width is 2 \kms.
Also obtained are the mean, its error, and the standard deviation of each distribution,
which are given in Table 2. 
An examination of the
figure and the table reveals that the histograms of Doppler velocities of the 
cell interior are broader than that of the network except in the case of He line
in the coronal hole.
The histograms of Doppler velocities of the network and cell interior
do not show significant difference in most cases. However, in the case of the 
quiet Sun, the Doppler velocities of the cell interior are more blueshifted than 
that of the network for the lowermost line He {\sc ii} 304 {\AA}, and an opposite
behavior is seen for the uppermost line Mg {\sc ix} 368 {\AA}
where it is the velocities of 
the network that are more blueshifted than that of the cell interior. 
For the intermediate lines from the quiet Sun, as well as for the lines from the 
coronal hole, the behavior is inconclusive because of the large uncertainties.

Histograms of linewidths are given in Figure~2.  The bin width is 0.01 {\AA}. 
The mean, error, and 
standard deviation are given in Table 3. It can be seen that the histograms of 
the cell interior are broader than those of the network in most of the cases. 
The exceptions are the O {\sc v} and Mg {\sc ix} lines in the quiet
Sun and the He {\sc ii} line in the coronal hole.
Also the network--cell interior difference is more prominent in the coronal hole.
The network has significantly larger linewidth than the cell interior for the lowermost TR line 
He {\sc ii} 304 {\AA} for the quiet Sun. For a coronal hole, this is true 
for the three lower TR lines He {\sc ii} 304 {\AA}, O {\sc iii} 599 {\AA}, 
and O {\sc v} 630 {\AA}. This shows that the network--cell interior difference disappears
faster in the quiet Sun.

The correlations between Doppler velocity and linewidth for the individual points
in the different regions are shown in Figure 3. A straight-line fit is given to show 
the overall behavior. Also given are the correlation coefficient and the probability
that the correlation can arise from two random distributions.

The correlation coefficients obtained from Figure 3 are plotted against the
formation temperatures of different lines
in Figure 4. A mild positive correlation is found for the lowest TR line 
He {\sc ii} 304 {\AA}. The correlation reduces or become
insignificant for the lower transition region lines O {\sc iii} 599 {\AA}
and O {\sc v} 630 {\AA}. The correlation is completely insignificant for the upper 
TR line Ne {\sc vi} 562.8 {\AA}. For the low coronal 
line  Mg {\sc ix} 368 {\AA}, the correlation becomes strongly negative.
It may also be seen that the correlation coefficients 
do not show any significant difference between the different regions.

  \begin{figure}    %%%%%%%%%%%%%%%%%% FIGURE 1 
   \centerline{\includegraphics[height=20cm,width=\textwidth,clip=]
              {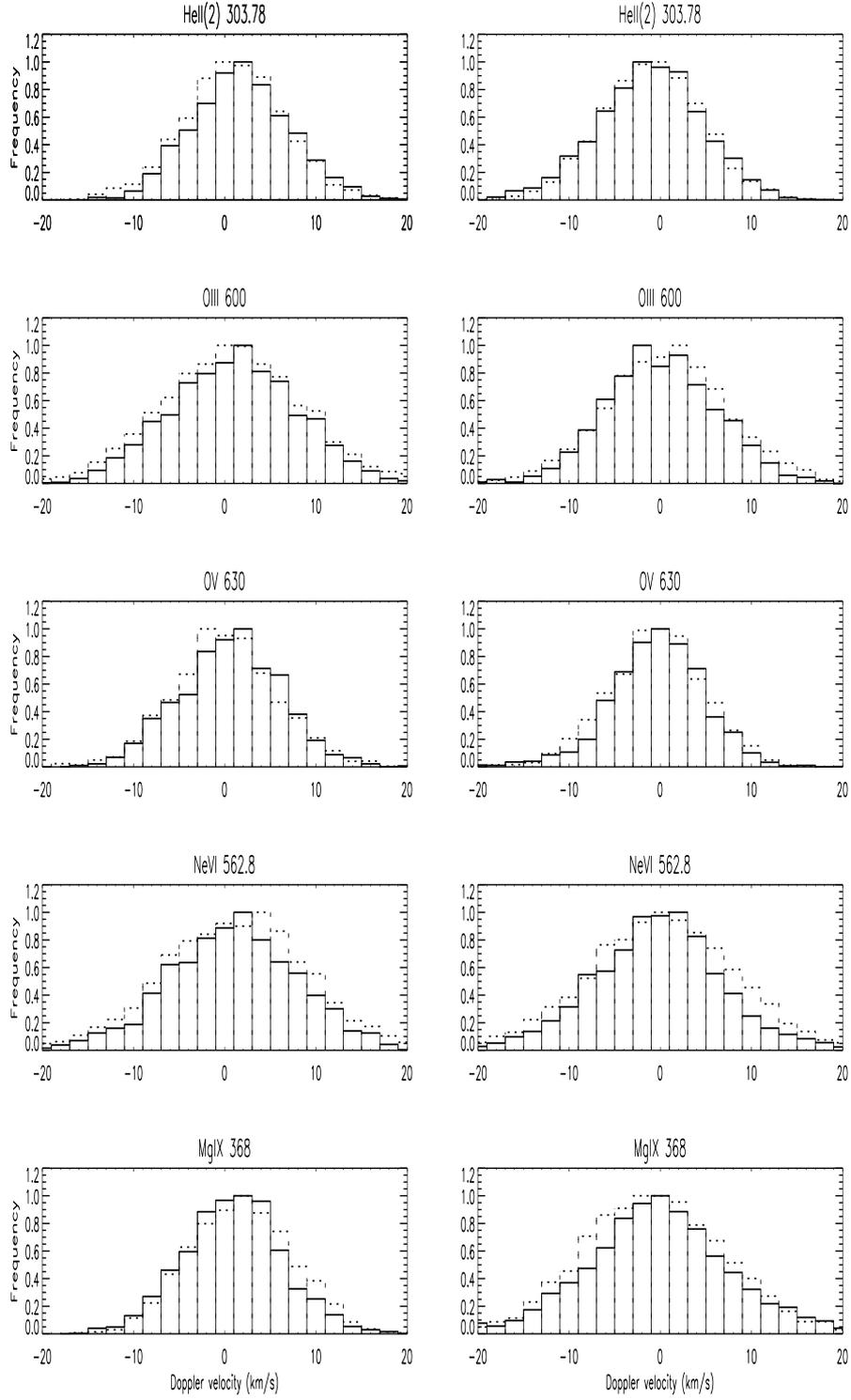}
              }
              \caption{Histograms of Doppler velocities for 
              different emission lines. The left panels represent quiet Sun and 
              right panels represent coronal hole. Solid line represents the 
              network and the dotted line represents cell interior.
              }
   \label{F-fig1}
   \end{figure}

\begin{table}[h]
\caption{ Details of velocity histograms (given in Figure~1). Mean of velocity 
distribution [vel] in \kms, its error, and standard deviation [stdv] of the network and cell
interior in the quiet Sun and coronal hole are given in different columns.
}
\label{T-table2}
\begin{tabular}{ccccccccc}
  \hline
No. & \multicolumn{4}{c}{QS}& \multicolumn{4}{c}{CH}\\
     & \multicolumn{2}{c}{network} & \multicolumn{2}{c}{cell} & 
       \multicolumn{2}{c}{network} & \multicolumn{2}{c}{cell}\\
     &vel&stdv&vel&stdv&vel&stdv&vel&stdv\\
  \hline
1&1.65$\pm$0.30&4.95&0.99$\pm$0.23&5.12&-1.02$\pm$0.69&5.29&-0.90$\pm$0.41&5.14\\
2&0.94$\pm$0.37&6.07&0.59$\pm$0.29&6.38&0.08$\pm$0.73&5.57&0.59$\pm$0.47&5.86\\
3&0.54$\pm$0.31&5.17&0.04$\pm$0.24&5.25&-0.27$\pm$0.62&4.72&-0.43$\pm$0.40&4.96\\
4&0.80$\pm$0.37&6.08&0.88$\pm$0.29&6.48&-0.32$\pm$0.78&5.98&0.08$\pm$0.53&6.56\\
5&0.83$\pm$0.31&5.05&1.59$\pm$0.24&5.19&-0.17$\pm$0.83&6.32&-0.59$\pm$0.52&6.43\\
  \hline
\end{tabular}
\end{table}

 \begin{figure}    %%%%%%%%%%%%%%%%%% FIGURE 2 
   \centerline{\includegraphics[height=20cm,width=\textwidth,clip=]
              {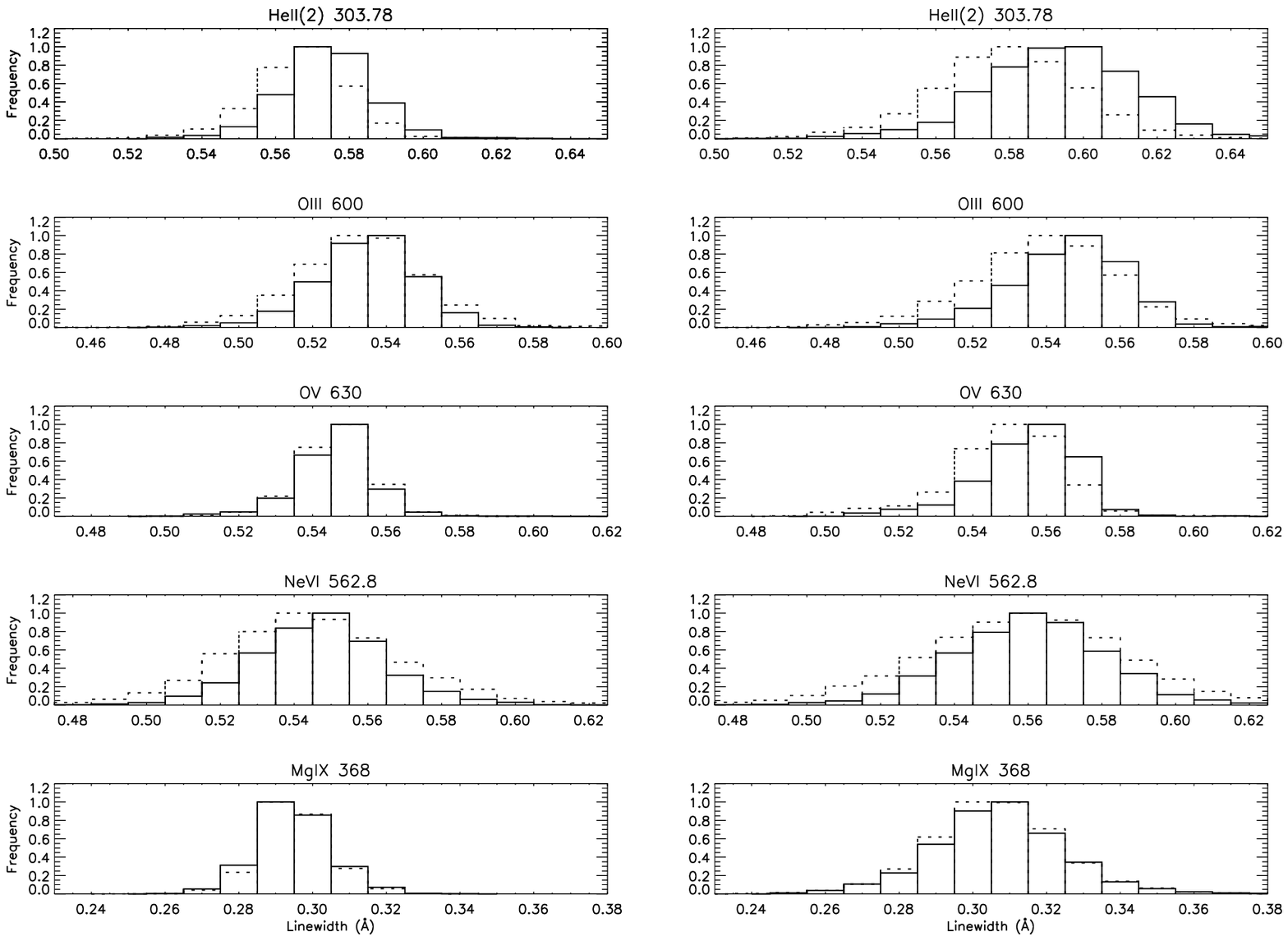}
              }
              \caption{Histograms of linewidths  for
              different emission lines. The left panels represent quiet Sun and 
              right panels represent coronal hole. Solid line represents the 
              network and the dotted line represents cell interior.
                      }
   \label{F-fig2}
   \end{figure}

\begin{table}[h]
\caption{ Details of linewidth histograms (given in Figure~2). Mean of linewidth 
distribution [wid] in {\AA}, its error, and standard deviation [stdv] of the network and cell
interior in the quiet Sun and coronal hole are given in different columns.
}
\label{T-table3}
\begin{tabular}{ccccccccc}
  \hline
No. & \multicolumn{4}{c}{QS}& \multicolumn{4}{c}{CH}\\
     & \multicolumn{2}{c}{network} & \multicolumn{2}{c}{cell} & 
       \multicolumn{2}{c}{network} & \multicolumn{2}{c}{cell}\\
     &wid&stdv&wid&stdv&wid&stdv&wid&stdv\\
  \hline
1&0.573$\pm$0.001&0.012&0.566$\pm$0.001&0.013&0.593$\pm$0.003&0.020&0.579$\pm$0.002&0.019\\
2&0.534$\pm$0.001&0.013&0.533$\pm$0.001&0.017&0.545$\pm$0.002&0.015&0.539$\pm$0.002&0.019\\
3&0.546$\pm$0.001&0.010&0.546$\pm$0.001&0.009&0.555$\pm$0.001&0.013&0.548$\pm$0.001&0.015\\
4&0.547$\pm$0.001&0.018&0.545$\pm$0.001&0.024&0.560$\pm$0.003&0.020&0.558$\pm$0.002&0.028\\
5&0.294$\pm$0.001&0.010&0.295$\pm$0.001&0.009&0.307$\pm$0.002&0.017&0.306$\pm$0.001&0.018\\
  \hline
\end{tabular}
\end{table}

 \begin{figure}    %%%%%%%%%%%%%%%%%% FIGURE 3
   \centerline{\includegraphics[height=20cm,width=\textwidth,clip=]
              {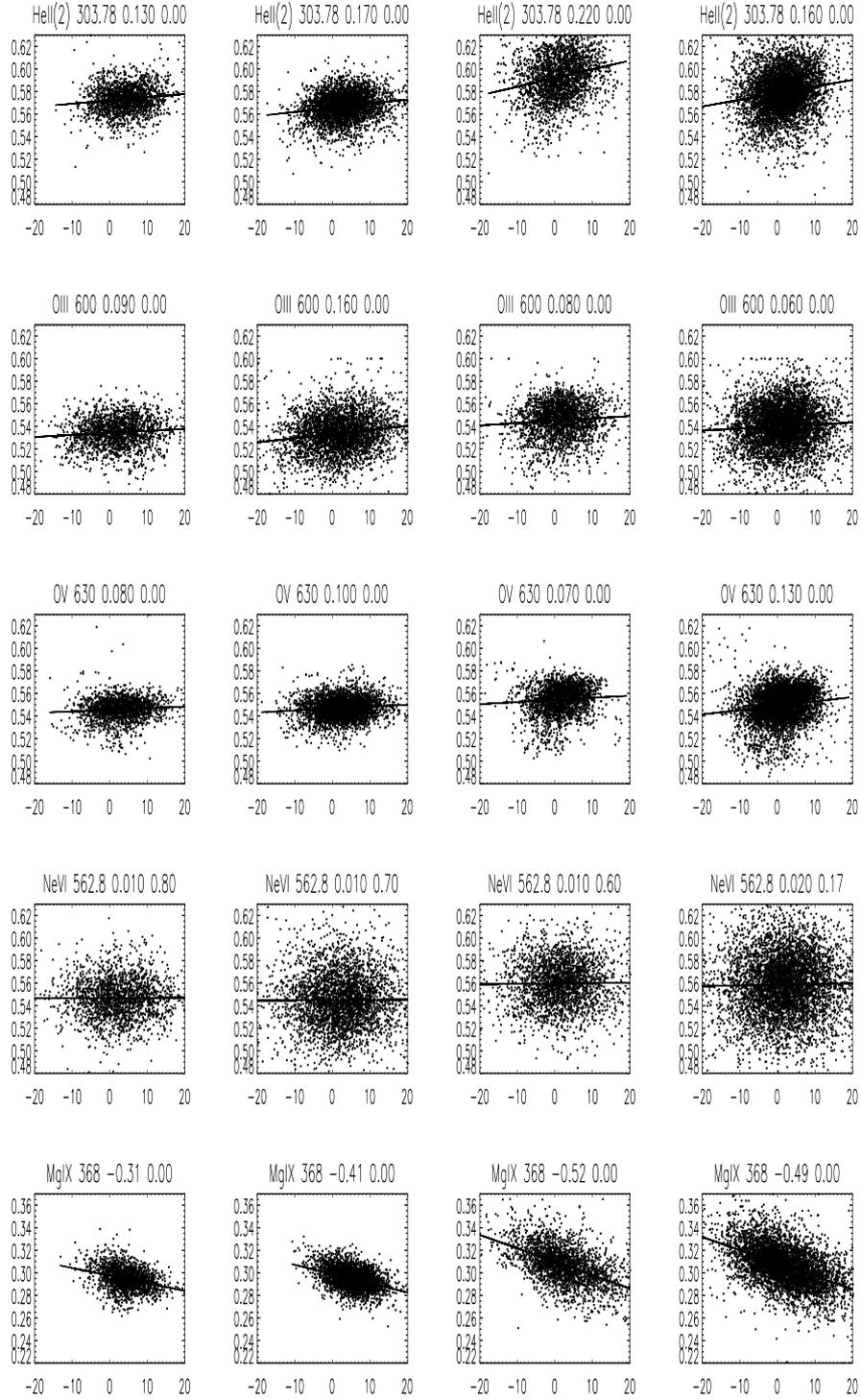}
              }
              \caption{Linewidths [\AA] are plotted against Doppler 
               velocities [\kms] for different emission lines. The line details,
               correlation coefficient and the probability that the correlation can 
               arise from two random distributions are given above each panel.
               From left to right; First panel represents network
              in quiet Sun, second represents cell interior in quiet Sun, third 
              represents network
              in coronal hole, and fourth represents cell interior in coronal hole. 
               }
   \label{F-fig3}
   \end{figure}

 \begin{figure}    %%%%%%%%%%%%%%%%%% FIGURE 4
   \centerline{\includegraphics[height=10cm,width=1\textwidth,clip=]
              {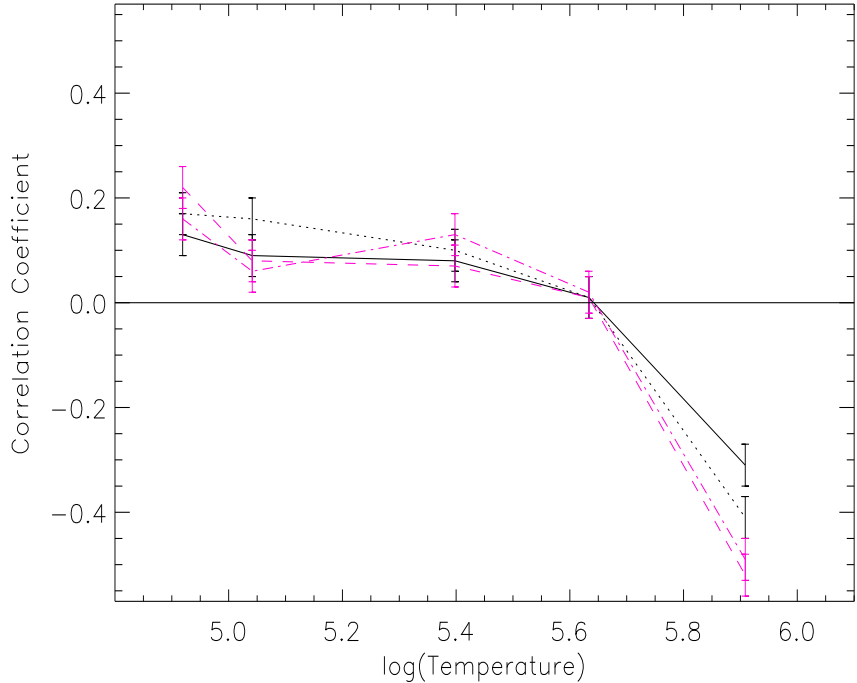}
              }
              \caption{Correlation coefficients plotted against log of formation 
              temperatures of the different lines. The solid line represents network in 
              quiet Sun, dots represent cell interior in quiet Sun, dashes represent network
              in coronal hole, and dash--dots represent cell interior in coronal hole. 
               }
   \label{F-fig4}
   \end{figure}

\section{Discussion}
\label{S-Discussion} 

Observations in recent years have shown that a coronal hole has larger blueshifts 
and linewidths as compared to a quiet-Sun region \cite{Ban98,Raju00,Stuc00}.
This has been understood as due to the origin of fast solar wind in the 
coronal hole \cite{Harra12}. The network and cell interior show large difference in their 
intensities \cite{Reev74, Gall98,Raju10} and widths (Figure~2) and therefore we
expect large differences in their velocities, which is not seen.
However, the pattern of difference seen in the lower-TR line 
He {\sc ii} 304 {\AA} and the low-coronal line Mg {\sc ix} 368 {\AA} in the quiet Sun
is interesting and is being reported for the first time. 
This also poses challenges for interpretation.
\inlinecite{Judge97} and \inlinecite{Gont01} report that the lower-TR lines show more 
redshift in the network than in the internetwork in the quiet Sun, which agrees
with our result. \inlinecite{Hass99} observe large blueshifts in the network in
coronal hole in the Ne {\sc viii} 770 {\AA} line which is formed at the base of 
the solar corona. Our results on the low coronal line Mg {\sc ix} 368 {\AA} 
from the coronal hole are inconclusive.
Before going further, these results need to be 
verified. 

The larger linewidths in the network as compared to the cell interior {\bf are} expected 
because the network is hotter than the latter. The results also show that the 
evolution of the network is faster in the quiet Sun than in the coronal hole. 
This is expected to be due to the difference in the thickness of the TR which is five 
times large in the coronal hole \cite{Huber74}. 

Plumes are known to have smaller linewidths and velocities compared to interplumes
\cite{Wil98}.
The excess width in the network of the coronal hole raises some questions 
regarding the network-origin of plumes. If plumes are indeed the extensions of the
network, then we might expect lower widths in the network. The results on the 
velocity distributions are, however, inconclusive.

The change of sign of the correlation coefficients of linewidth and velocity for
lines from transition region to the corona is being reported for 
the first time, although some evidence of this can be seen by  \inlinecite{Xia04}.
In their work with {\it Solar Ultraviolet Measurements of Emitted Radiation} (SUMER) 
data, the Si {\sc ii} 1533 {\AA} line shows a positive correlation
while the O {\sc vi} 1038 {\AA} line shows a negative correlation and the behavior of other 
lines is ambiguous.
The behavior of correlation coefficients could be due 
to the presence of standing or propagating waves
from the lower to the upper solar atmosphere. For example, Alfv\`{e}n 
waves can cause simultaneous variations in width and velocity. 
Alfv\`{e}n waves are known to be one of the candidates for coronal heating.

\section{Conclusions}
\label{S-Conclusions} 

We have obtained the Doppler  velocities and emission linewidths in a coronal hole 
and the nearby quiet Sun in five different emission lines originating at different
heights in the solar atmosphere from  the  lower TR to the inner corona. The behavior
of the velocities and widths in the network and the cell interior in the coronal hole
and quiet Sun were examined. 
Histograms of Doppler velocity and width are 
generally broader in the cell interior as compared to the network. 
The histograms of Doppler velocities of the network and cell interior
do not show significant differences in most cases with exceptions in 
the lower-TR line He {\sc ii} 304 {\AA} and the low-coronal line Mg {\sc ix} 368 {\AA}
in the quiet Sun. Doppler velocities of the cell interior are more blueshifted than 
that of the network for the He line and  an opposite
behavior is seen for the Mg line.
The histograms of line width show that the network--cell difference is 
more prominent in the coronal hole and the network--cell merger happens faster in the quiet Sun.
A mild positive correlation between the relative Doppler velocity and the 
linewidth is found for the lowermost transition region line He {\sc ii} 304 {\AA},
which further reduces or become insignificant for the intermediate lines, and becomes 
strongly negative for the low coronal line,  Mg {\sc ix} 368 {\AA}.
This could be due to the 
presence of standing or propagating waves from the lower to the upper solar atmosphere

\begin{acks}
Data are provided courtesy of SOHO/CDS consortium. SOHO is a project of 
international cooperation between ESA and NASA.
\end{acks}

\end{article} 

\end{document}